# Two-Center Gaussian potential well for studying light nucleus in cluster structure

Nafiseh Roshanbakht[1], Mohammad Reza Shojaei[1]

1.Department of physics, Shahrood University of Technology

P.O. Box 36155-316, Shahrood, Iran

Shojaei.ph@gmail.com

## Abstract

The clustering phenomena is very important to determine structure of light nuclei and deformation of spherical shape is inevitable. Hence, we calculated the energy levels of two-center Gaussian potential well including spin-orbit coupling by solving the Schrödinger equation in the cylindrical coordinates. This model could predict the spin and parity of the light nucleus that have cluster structure consist of two identical clusters.

**Key words:** energy levels, Gaussian potential, spin-orbit coupling, cluster structure

## 1. Introduction

In the light nucleus, deformation plays an important role in determining nuclear structure. This deviations from spherical structures are found in the axial deformations and the clustering because numerous experimental studies have revealed a clustering phenomena in them [1], [2]. Freer and Merchant in 1997, studied the role of clustering and cluster models in nuclear reactions and examined the evidence for α-cluster chain configurations in the light even - even nuclei from $^8$Be to $^{28}$Si [3], [4]. In 2001, Kanada and Horiuchi showed that the valence neutrons in nucleus with cluster structure can form a covalent bond between the clusters and named it "nuclear molecules"[5]. For example $^9$Be has a neutron that exchange between the two alpha clusters. They studied them by the one-center deformed harmonic oscillator potential[5], [6].

For studying light nucleus in cluster model, many physicists used various methods such as: Antisymmetrized molecular dynamics (AMD) [7] and Fermion molecular dynamics (FMD) [8]and condensates and the THSR Wave-Function [9]–[11]. These models have many important advantages for studying nuclear in cluster models, but don't make assumptions about the cluster [12].

The two-center shell model is more appropriate for modeling nuclei with clustered nature. This model assumes a clustered structure, and each cluster is represented by its own potential. This potential is the two-center harmonic oscillator (TCHO)[13]. It has a complete analytical solution but does not have a sufficiently sharp edge to satisfy finite separation energies.

In this article, we consider an axially symmetric nuclear parameterization, so the nuclear surface equation is given in the cylindrical coordinates [14], [15]. In this case, the deformation removes the degeneracy of single nucleon levels associated with spherical symmetry [16]. In the next section, we introduce new potential for studying the nucleus with the cluster structure.

## 2. 2D potential in cylindrical coordinates

If the nucleus structure consists of two identical clusters, the nuclear surface in the cylindrical coordinates is defined by [15], [17]:

$$\rho = \begin{cases} b\sqrt{1-\left(\dfrac{z+c_1}{a}\right)^2} & z \leq z_1 \\ \rho_2 - \sqrt{R^2 - (z-c_2)^2} & z_1 \prec z \prec z_2 \\ b\sqrt{1-\left(\dfrac{z-c_1}{a}\right)^2} & z_2 \leq z \end{cases} \qquad (1),$$

where $\pm c_1$ are the positions of the two cluster centers and other parameters are shown in Fig.1.

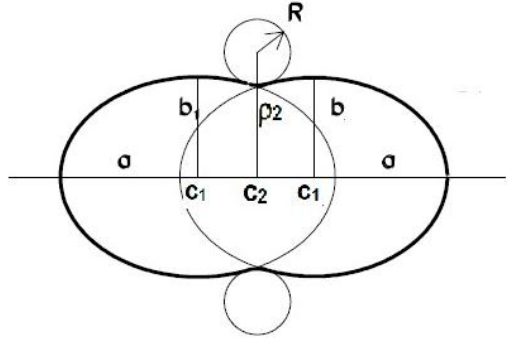

**Fig.1: Parameterizations of nuclear shape in two cluster structure**

Hence, we use two-center Gaussian potential well for calculating the energy levels of nucleus that are nearly spherical or deformed because of having cluster structure. Our potential in cylindrical coordinates can be split into two parts:

$$V_0(\rho,z) = \begin{cases} A\exp(-\alpha(z-z_0)^2) + B\exp(-\beta\rho^2), & z \geq 0 \\ C\exp(-\lambda(z+z_0)^2) + B\exp(-\beta\rho^2), & z < 0 \end{cases} \qquad (2),$$

where A,C and B are the potential depth, the parameters $\alpha,\beta$ and $\lambda$ are inverse square range and $z_0$ is cluster centers. We study nucleus that have two identical cluster structure. Hence, we suppose the parameters A =C and $\alpha=\lambda$ to have a symmetric potential form. In Fig.2, the potential is plotted for some valuable of $z_0$. The nuclei with the closed shell such as $^4$He nucleus have a spherical shape and are exceptionally stable. Hence, the spherical Gaussian potential well (Fig. 2a) can describe their energy levels. But in light nucleus with almost half full shells N=2, the nucleons have been observed to cluster structure with a two-center structure. For example $^8$Be isotope at the ground state has two alpha cluster structure and decays to alpha particles by 92 keV energy. In this case, the potential in Fig. 2d can be used describe it. Likewise, for light neutron-rich nucleus that have covalent binding between two clusters due to valence neutrons, the potentials are shown in Fig.2b and Fig. 2c.

If the potential of the quantum system to be examined is axially symmetric, then the Schrödinger equation in the cylindrical coordinates can be employed. For a three-dimensional problem, the Laplacian in cylindrical coordinates is used to express the Schrödinger equation in the form bellow:

$$\frac{-\hbar^2}{2M}\nabla^2\psi(\rho,\phi,z)+V_0(\rho,z)\psi(\rho,\phi,z)=E\psi(\rho,\phi,z) \quad (3),$$

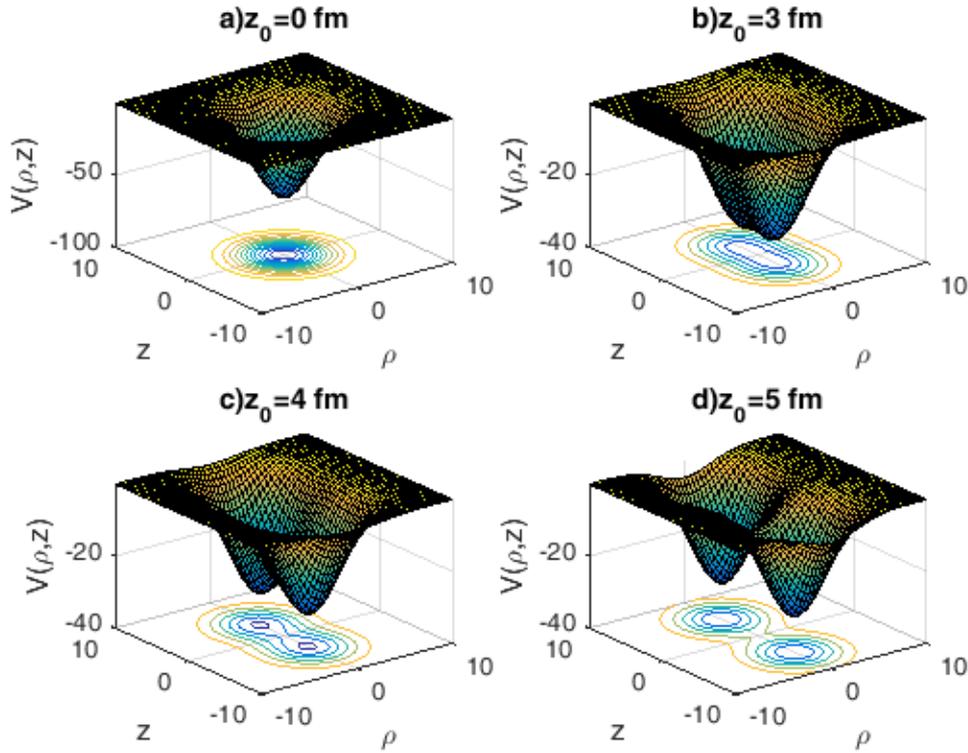

Fig.2: The Gaussian potential for some valuable of $z_0$

where $E$ and $M$ denote the energy eigenvalue and the mass, respectively. $V_0(\rho,z)$ is potential that does not depend on $\phi$ because of the axially symmetric and is separable to $V_0(\rho)$ and $V_0(z)$. In the cylindrical coordinates, the Laplace operator takes the form:

$$\nabla^2=\frac{\partial^2\psi}{\partial\rho^2}+\frac{1}{\rho}\frac{\partial\psi}{\partial\rho}+\frac{1}{\rho^2}\frac{\partial^2\psi}{\partial\phi^2}+\frac{\partial^2\psi}{\partial z^2} \quad (4).$$

The separation of variables is accomplished by substituting:
$$\psi(\rho,\phi,z)=R(\rho)Z(z)\Phi(\phi) \quad (5).$$
In the usual way, this leads to the following ordinary differential equations[18]:

$$\frac{d^2\Phi(\phi)}{d\phi^2}+m^2\Phi(\phi)=0 \quad (6),$$

$$\frac{d^2Z(z)}{dz^2}-\frac{2M}{\hbar^2}(V_0(z)-E)Z(z)+\gamma^2 Z(z)=0 \quad (7),$$

$$\frac{d^2R(\rho)}{d\rho^2}+\frac{1}{\rho}\frac{dR(\rho)}{d\rho}-(\gamma^2+\frac{m^2}{\rho^2})R(\rho)-\frac{2M}{\hbar^2}V_0(\rho)R(\rho)=0 \quad (8).$$

The solution of the first equation is:

$$\Phi(\phi) = \frac{1}{\sqrt{2\pi}} \exp(im\phi) \tag{6},$$

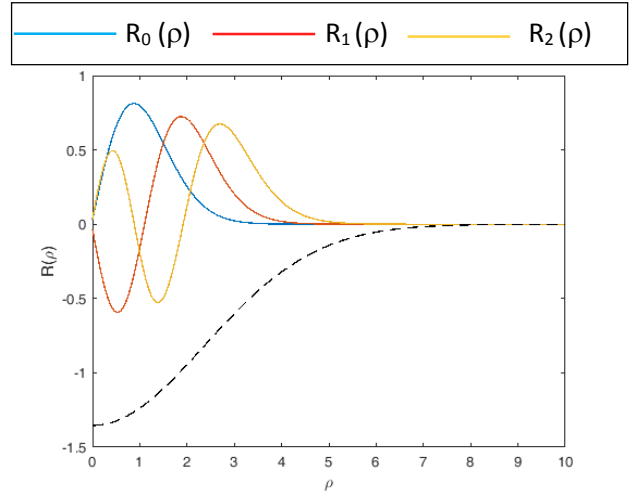

**Fig.3: The eigenvectors $R_{n_\rho}(\rho)$ of three lowest states**

where *m* is the magnetic quantum number and integer. The next step is to obtain the solutions of Eq. 7 and Eq. 8 that cannot be solved by analytical methods. But, the Hamiltonian is diagonal, and there are some numerical method to solve it. We use a well-known Numerov numerical methods to solve ordinary differential equations of second order[19], [20]. In this method, we first organize the square matrix Hamiltonian of the system consist of the kinetic energy matrix and the potential energy matrix, then we obtain the eigenvectors of Hamiltonian and the eigenvalues for the stationary states of the time-independent Schrödinger equation. The values of M and ℏ are considered M=1 and ℏ=1. The eigenvectors, $R_{n_\rho}(\rho)$ for three lowest states are shown in Fig. 3 and the eigenvectors, $Z_{n_v}(z)$ are plotted for some values of $z_0$ in Fig. 4. $n_\rho$ is the radial quantum number and $n_z$ is the quantum number along *z* axis having values 0, 1, 2, … .

The eigenvectors $R_{n_\rho}(\rho)$ vanish on the boundaries of our system and behave as expected. For the eigenvectors $Z_{n_v}(z)$, in addition to boundary conditions, we consider the continuity conditions too:

$$\underset{z<0}{Z_{n_v}(z \to 0)} = \underset{z>0}{Z_{n_v}(z \to 0)} \tag{7},$$

$$\underset{z<0}{\frac{\partial Z_{n_v}(z \to 0)}{\partial z}} = \underset{z>0}{\frac{\partial Z_{n_v}(z \to 0)}{\partial z}} \tag{8}.$$

While we describe a nuclei in an axially symmetric model, the orbital angular momentum, *l*, and the intrinsic spin, *s*, are not good quantum numbers because states with different *l*-values with the same parity can mix. In this model, the quantum number $N=n_\rho+n_z$ is the principal quantum number and the parity is determined by $(-1)^N$. The magnetic quantum number having the integer values between $-n_\rho$ to $n_\rho$ too.

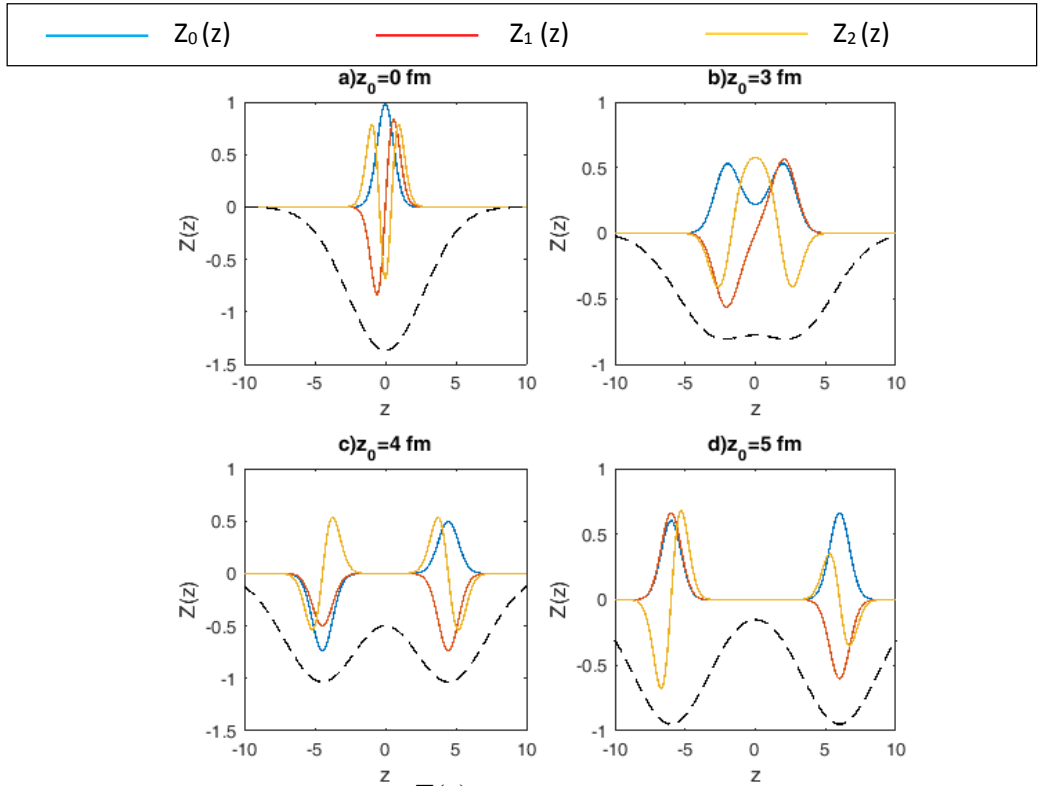

**Fig.4: The eigenvectors $Z(z)$ of three lowest states for some values of $z_0$**

The Figs. 4a to 4d show that the wave functions associated to z axis have a definite parity with respect to z=0. For finite separations the wave functions do not feel the presence of the other cluster across the barrier in low energy. The energy levels with the same quantum number that located in the first or in the second potential well have the same value.

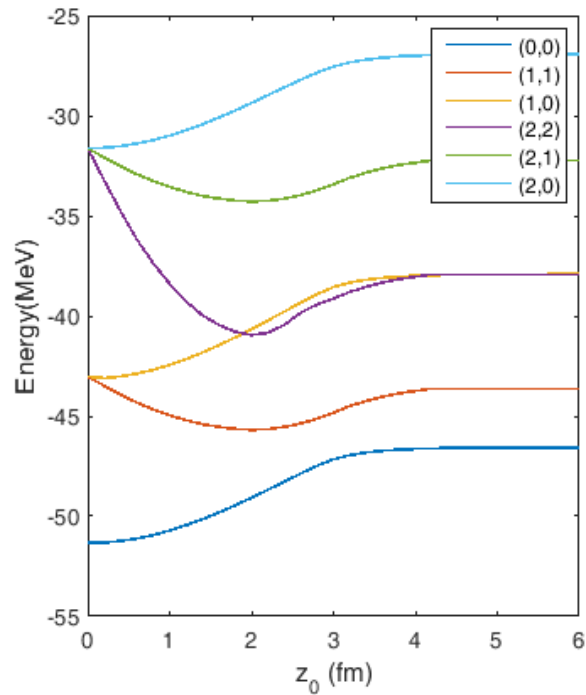

**Fig. 5: Variations of the eigenvalues in z axis with the distance between the two cluster centers**

The eigenvalues from the solution of the Hamiltonian equation in z axis is obtained for the different values of the distance between the two cluster centers, and the variation of them is given in Fig. 5. In this Figure, the shell structure of the spherical nucleus is reproduced for the two separate clusters, in higher energy. At $z_0=0$ there are the principal levels without spin-orbit coupling (N, m) and for large values of $z_0$, the levels start to form two clusters and their levels overlap each other.

One of the interesting feature of Fig.5 is that states having the same z-component of angular momentum, m, repulse each other, and prevent crossing. This is consistent with the theory Neumann and Wigner. They showed such level crossings are highly improbable [21].

The levels with different values of $n_\rho$ and $n_z$, but the same value of $N$ are degenerate. For example *($n_z$, $n_\rho$, m)=(1,0,0)* are degenerate with *($n_z$, $n_\rho$, m)=(0,1,0)*. We calculated the spin-orbit coupling to eliminate this degeneracy.

## 3. The effect of spin-orbit coupling on levels

After the presentation of the solutions of the Schrödinger equation with the potential in Eq. 2, we want to explain the effect of spin-orbit coupling on the single particle levels for a spherical nuclei and for a nuclei with two cluster structure separated at an infinite distance.

The spin-orbit coupling is defined in the form of [15], [22]:

$$V_{L.S} = -\lambda \left(\frac{1}{2m_0 c}\right)^2 \vec{L}\vec{S} \qquad (9),$$

where $\lambda = 35$ is a dimensionless coupling constant, $m_0$ and c are the nucleon and the speed of the light respectively[15]. The orbital angular-momentum operators is $\vec{L} = \vec{\nabla}V \times \vec{p}$ and we must transform it into cylindrical coordinates:

$$\hat{L}^\pm = \mp \hbar e^{\pm i\varphi}\left(\frac{\partial V}{\partial \rho}\frac{\partial}{\partial z} - \frac{\partial V}{\partial z}\frac{\partial}{\partial \rho} \pm i \frac{\partial V}{\partial z}\frac{1}{\rho}\frac{\partial}{\partial \varphi}\right) \qquad (10).$$

$$\hat{L}_z = i\hbar \frac{\partial V}{\partial \rho}\frac{1}{\rho}\frac{\partial}{\partial \varphi} \qquad (11),$$

hence:

$$\vec{L}\vec{S} = \frac{1}{2}(L^+ s^- + L^- s^+) + L_z s_z. \qquad (12).$$

The shift of the energy levels due to the spin-orbit coupling was calculated using the orthogonality of the wave functions. The matrix elements of the spin operators are:

$$\langle m'_s |\hat{s}_+| m_s\rangle = \hbar \delta_{m'_s, m_s+1} \quad , \langle m'_s |\hat{s}_-| m_s\rangle = \hbar \delta_{m'_s, m_s-1} \quad , \langle m'_s |\hat{s}_z| m_s\rangle = \hbar m_s \delta_{m'_s, m_s} \qquad (13)$$

and the spin-orbit matrix is:

$$\langle n'_z, n'_\rho, m', m'_s | \frac{1}{2}(L^+s^- + L^-s^+) + L_z s_z | n_z, n_\rho, m, m_s \rangle = \frac{\hbar}{2} \langle n'_z, n'_\rho, m' | L^+ | n_z, n_\rho, m \rangle \delta_{m'_s, m_{s-1}}$$

$$+ \frac{\hbar}{2} \langle n'_z, n'_\rho, m' | L^- | n_z, n_\rho, m \rangle \delta_{m'_s, m_{s+1}} \quad (14).$$

$$+ \hbar m_s \langle n'_z, n'_\rho, m' | L_z | n_z, n_\rho, m \rangle \delta_{m'_s, m_s}$$

Now, we use the parabolic approximation [23], and consequently the levels of single particle energies are obtained as a function of $z_0$ by the diagonalization of the potential $V_0+V_{L.S}$. The structure of levels is plotted in Fig. 6 and labeled them by $j_z$. In our model, the projection of the total angular momentum on the symmetry axis, $j_z = m+m_s$, is the good quantum number. Hence, states of different $j_z$ are not coupled by the Hamiltonian but the states with $\pm j_z$ are degenerate because the nuclei have reflection symmetry for either of the two possible directions of the symmetry axis.

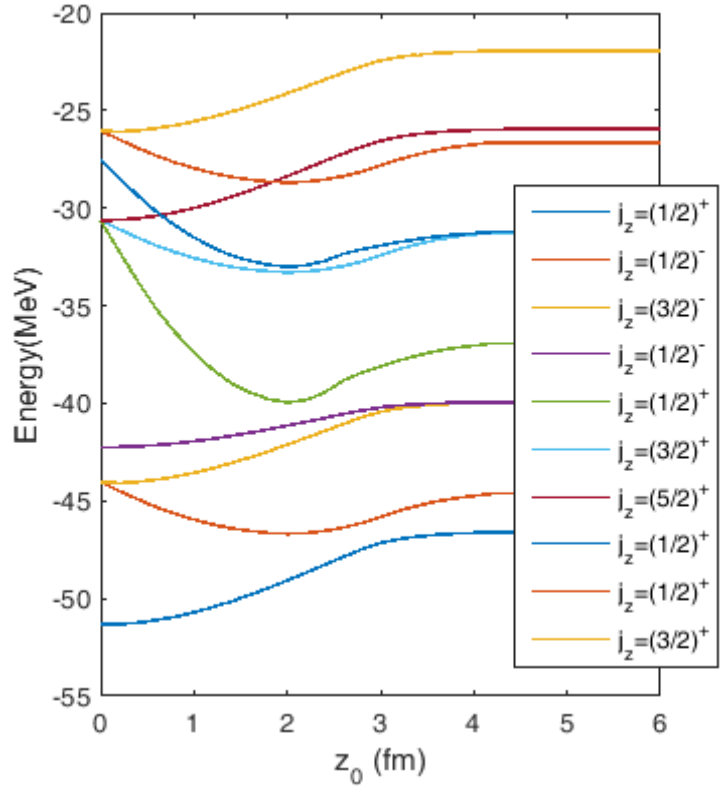

**Fig. 6: Variations of the eigenvalues ($n_z, n_\rho, m$) in z axis including spin-orbit coupling with the distance between the two cluster centers**

In Fig. 6, the level scheme of the spherical nuclei is obtained for $z_0=0$. But by increasing $z_0$, two separated clusters are formed. An example of a light nuclei by two cluster structure is $^9$Be isotope. The nucleon single-particle level scheme is plotted in Fig. 6. $^9$Be isotope is stable and have 4 protons and 5 neutrons. Alpha cluster is in the first level and in the second level will be other alpha cluster and the remaining single nucleon located in the next level. Experimental evidence for $^9$Be shows it has a ground state with spin and parity 3/2⁻ and a 1/2⁺ excited state[24], [25]. For predicting these results, we must use the potential well in Fig. 4b. When two separate center in potential well is formed, the level (1,0,0) and (0,1,0) overlap with each other. In our model, the single neutron is in the level $J_z=3/2$ at the ground state and its parity is $(-1)^N=-1$. For the first excited state, the single neutron move into level $J_z=1/2^+$. We will predict other excited state by considering the rotational band and Coulomb potential between the clusters.

## 4. Conclusion

In light nuclei, close to decay threshold, the cluster phenomenon is the favored mode. Hence, we use two-center Gaussian potential well for calculating the energy levels of nucleus that are nearly spherical or deformed because of having two identical cluster structure. We solved the Schrödinger equation in the cylindrical coordinates, and plotted the level scheme without considering spin-orbit coupling. Then we modified results with it. For $z_0=0$, the level scheme of the spherical nuclei is obtained, and two separated clusters are formed by increasing $z_0$. In principle, a two-center potential well can be used to study light nuclei with two cluster structure and make assumptions about the cluster in them.